# A MEMS-scale vibration energy harvester based on coupled component structure and bi-stable states

Masoud Derakhshani, Ph.D. Candidate, University of Louisville, 200 Sacket Hall, Louisville, KY 40208
Brian E. Allgeier, Master Student, University of Louisville, 200 Sacket Hall, Louisville, KY 40208
Thomas A. Berfield, Associate Professor, University of Louisville, 200 Sacket Hall, Louisville, KY40208

**ABSTRACT**

Due to the rapid growth in demand for power for sensing devices located in remote locations, scientists' attention has been drawn to vibration energy harvesting as an alternative to batteries. As a result of over two decades of micro-scale vibration energy harvester research, the use of mechanical nonlinearity in the dynamic behavior of the piezoelectric power generating structures had been recognized as one of the promising solutions to the challenges presented by chaotic, low-frequency vibration sources found in common application environments. In this study, the design and performance of a unique MEMS-scale nonlinear vibration energy harvester based on coupled component structures and bi-stable states are investigated. The coupled-components within the device consist of a main buckled beam bonded with piezoelectric layers, a torsional rod, and two cantilever arms with tip masses at their ends. These arms are connected to the main beam through the torsional rod and are designed to help the main beam snap between its buckled stable states when subjected to sufficient vibration loading. The fabrication of the device will be discussed, including use of plasma-enhanced chemical vapor deposition (PECVD) of silicon nitride under an alternating power field to control compressive stress development within the main buckled beam. After completing the fabrication process, the next step would be testing the device under a variety of vibration loading conditions for its potential use as a vibration energy harvester.

**Introduction**

Vibration energy harvesting is a well-known method to harvest electrical energy from ambient mechanical motions. For linear systems, this method is firmly rooted in the typical equations of motion for mechanical vibration, which leads such systems to work efficiently only at a narrow bandwidth close to their natural frequencies. It was found that using physical properties of mechanical nonlinearity could be a good solution to this issue and significantly improve both the frequency bandwidth and the output power of energy harvesters.

There have been numerous studies on MEMS vibration energy harvesters to improve their functionality and make them work in real-life applications. H.B. Fang et al.[1] designed and fabricated a MEMS-scale piezoelectric-based vibration energy harvester using different techniques of micro-fabrication. Their experimental testing of the fabricated devices showed that improved microfabrication methods could produce good performance compared to typical micro-scale vibration energy harvesters. J.Q. Liu et al.[2] developed a MEMS-scale piezoelectric based power generator consisting of cantilevers array to improve the frequency range and power output. The structural material for the cantilever was silicon with PZT film as a transducer and a proof mass is attached to the free end of each beam. The experimental results showed that the arrayed design of cantilevers with different dimensions makes the system work at a broader range of frequencies and ultimately is able to generate more output power. M.Ferrari et al.[3] studied the efficiency enhancement of a nonlinear piezoelectric vibration energy harvester coupled with a permanent magnet. A piezoelectric bimorph cantilever was considered as the harvester. The tip mass of the cantilever was faced towards a fixed permanent magnet, which serves a role in creating bi-stability in the system. The experimental results showed that for a low enough distance between the tip mass and the magnet, a softening behavior could be captured which shifts the resonances towards lower frequencies. It was also observed that snap-through motion can be created when the system is excited by white-noise mechanical vibrations. In 2015, M. Rezaeisaray et al.[4] designed and fabricated a multi-degree of freedom system as a vibration energy harvester functional at low excitation frequencies. In their proposed design, a big proof mass is connected to the device frame by two cantilever beams. The structure is fabricated from a silicon substrate, and aluminum nitride was used as the electromechanical converter material. One of the advantages pointed out in this work was being able to reduce the first three natural frequencies of the micro-scale size of the system to within the range of ambient vibrations. The potential of a flexible structure consisting of multiple beams connected with joint masses in a zigzag combination as a vibration energy harvester was studied by S. Zhou et al.[5]. In this study, a theoretical model was proposed and compared with both finite element analysis and experimental results. The obtained results showed that the resonant frequencies of the system can be significantly decreased by increasing the number of the considered beams and, consequently, can be more practical for low-frequency ambient vibrations. Another advantage was pointed out in this research is the functionality of this device under multi-directional excitations, which makes it a suitable approach to harvest energy from human body motion.

While a lot of studies have been done so far on the improvement of functionality of micro-scale vibration energy harvesters, the absence of a solid promising design to address the above-mentioned issues is still needed in this area. In this study, steps have been taken towards fabricating a unique-designed MEMS structure as a vibration energy harvester, which has a very high potential to resolve the literature mentioned problems and work efficiently at ambient vibration frequency levels.

**Model Description and Experimental Process**

A computer model of the proposed MEMS-scale device described in this research is shown in Fig 1, consisting of a main buckled beam, in which the buckling created by a compressive load. This beam is connected to two cantilever arms via a clamped-ends rod which role is transferring motion between the main buckled beam and these two arms. In order to tune the device flexibility and lowering its natural frequencies, a tip mass is considered at the free end of each arm. Different colors of the top view in this figure is related to different layers of the device including structural layer, bottom electrode, piezoelectric layer and top electrode. Dynamic behavior of a macro-scale version of this structure has been analyzed in [6] and it was shown that the dynamic performance of the structure is significantly dependent on different parameters including the level of compressive load as well as the geometry of different components and for certain cases, it is possible to capture snap-through motion, which is crucial for its functionality as a vibration energy harvester.

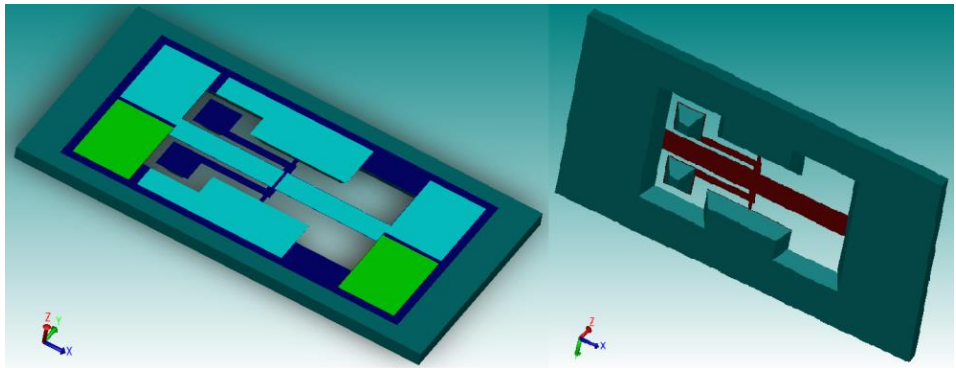

**Fig.1**. Schematic design of the proposed MEMS vibration energy harvester; left figure: top view; right figure: bottom view

The fabrication of the proposed device requires five photomasks corresponding to patterning of the following material layers or processes: stressed silicon nitride, bottom electrode, piezoelectric, top electrode, and the final bulk etching step. These masks are displayed in Fig 2 superimposed on top of one another. The main structural layer of the device is made of stressed nitride, on which the electrode and piezoelectric layers are deposited. All microfabrication processes discussed in this research have taken place at the University of Louisville's Micro/Nano Technology Center (MNTC) [7].

All fabrication processes of this device were performed on a four-inch silicon wafer. The Si wafer maintains a thickness of 380 µm and (100) crystalline orientation. The main structural layer considered for this device is stressed nitride. This material is the first fabricated layer supplies the support for the top layers of the device. Furthermore, the controlled stress created in this layer provides the required buckling load for the main beam. This layer is fabricated using plasma-enhanced chemical vapor deposition machine (PECVD) at the University of Louisville's Micro/Nano Technology Center.

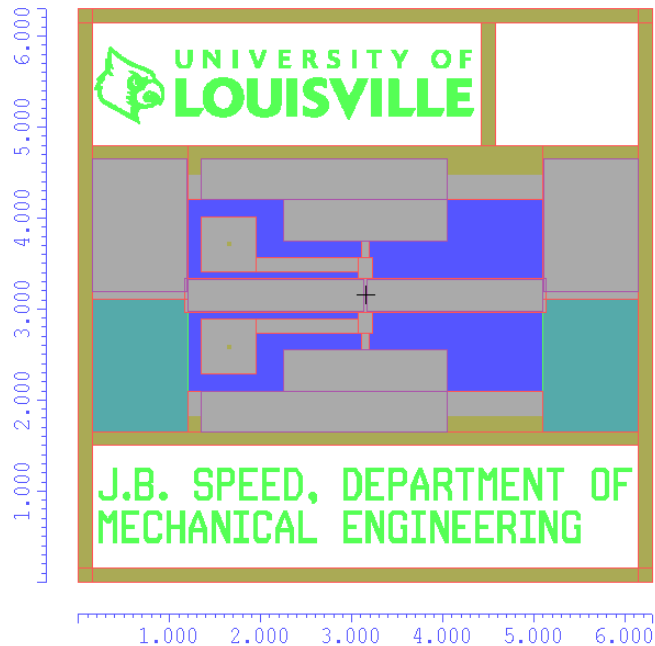

**Fig.2**. Superimposed layouts of all five device photomasks: bulk etching (blue), stressed nitride (green), bottom electrode (cyan), piezoelectric (red), top electrode (violet).

Molybdenum is considered as the metal material for both the bottom and top electrodes. These two layers are electrically connected to the top and bottom face of the piezoelectric layer, which role is generating electrical current from the induced strain in the main structure created by mechanical vibrations. All metal layers were deposited via physical vapor deposition (PVD) process and were patterned with their own masks in a photolithography process, followed by a dry etching step with a Trion Metal Etcher tool. The piezoelectric material is deposited and patterned between two electrode layers, which is responsible for harvesting electrical current from the vibration load. The material considered for this part is aluminum nitride. The deposition process was taken the same way as mentioned in [8] by using PVD sputtering machine. In this method, the target chosen for the sputtering process is aluminum which was combined with a certain portion of nitrogen and argon gasses. The last step of the fabrication process is backside etching, which releases the whole structure by etching away the bottom silicon layer of the wafer and turned out to be a little challenging due to the type of etching process had to be chosen for this specific device. Because of some certain required rules of the UofL Micro/Nano Technology Center, it has been decided to use anisotropic wet etching step for releasing the structure from its substrate. The etching technique was used in this research is potassium hydroxide (KOH) wet etching which etches straight walls at a 54.7° angle. A software named ACES (Anisotropic Crystalline Etching Simulator) was used to simulate the whole process of bulk etching and based on the software calculation, the required etchant solution for the considered wafer thickness and orientation is 45% of KOH.

**Results and discussion**
All the surface micro-fabrication processes discussed in the previous section were successfully performed step by step and a picture of the wafer after fabrication of the top electrode layer is shown in Fig 3. The piezoelectric effect of the deposited aluminum nitride layer was successfully tested by an oscilloscope and it was shown some noticeable spikes when the wafer sample was lightly struck, creating a small amount of strain.

The challenging part of the fabrication is the very last step, which releases the whole structure from its substrate and has been tried to be done by KOH wet etching technique. Unfortunately, the results obtained for stressed nitride layer were not the expected ones and the moment that the structural layer is released, the device could not be held together and bursts after the etching process. This phenomenon can be due to multiple reasons. One of them is the amount of induced stress in the structure which is high enough to break the whole top layers when they lose their support from the substrate. In order to check this possibility, a test was performed on a wafer with a patterned nitride layer via unstressed nitride recipe. The mentioned release process was applied on the test wafer to free the unstressed nitride layer. It was shown that the nitride layer could be successfully freed in many cells without any major damage (Fig 4).

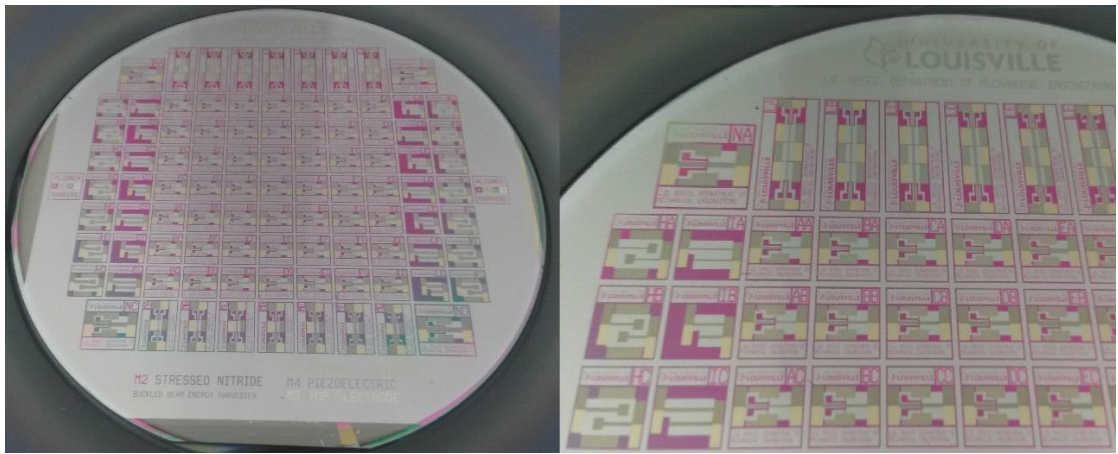

**Fig.3**. A processed wafer after the deposition and patterning of the tope electrode layer

Another reason may cause such thing to happen is the lack of rigidity of the structural layer due to not having enough thick stressed nitride layer on the top. To tackle this issue, it needs to increase the time deposition of stressed nitride via PECVD machine. However, increasing the thickness of the structural layer could lead to a less flexible buckled beam and consequently make it hard for the device to switch between its stable states as it is expected to do. One more thing needs to be considered in this study is the method chosen for releasing the structure from the backside substrate. Although the etching parameters were chosen based on our precise calculations from the simulation results, over-etching phenomenon, as well as the possibility of leaking KOH solution into the wafer fixture and destroying the unexposed side of the processed wafer, were two major problems created during the experimental process. One way to resolve this issue is using deep reactive-ion etching method (DRIE) instead of wet KOH etching, which can result in more controlled process and accurate geometries and consequently better output.

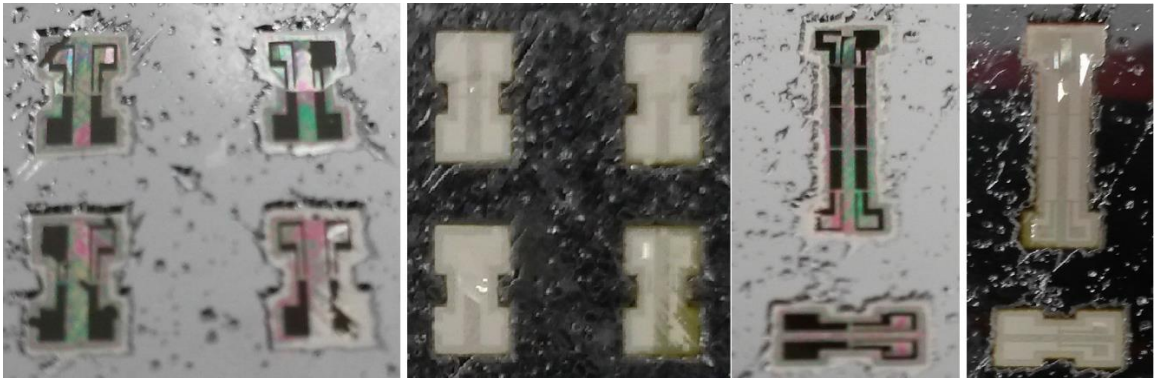

**Fig.4**. Several examples of successfully released devices for unstressed nitride layer

**Summary and conclusion**
In this study, design and fabrication of a micro-scale bi-stable vibration energy harvester consisting of different coupled components have been thoroughly investigated and the performed steps of each process were discussed in detail. As mentioned in the previous sections, it has been achieved very successful results for the surface fabrications of different layers of the device as expected. However, the very last step of bulk etching was failed to perform by KOH wet etching method, which made it impossible to dynamically test the device for its functionality. There are multiple reasons for this issue, including the large compressive stress created in the nitride layer and the lack of rigidity of the device due to too much thinness of the top deposited layers. While these mentioned issues can be resolved in the fabrication process by further experiments, there needs to be a balance between the main beam flexibility and the induced stress of the structural layer, so the output device is able to satisfy the expected effects of the desired bi-stable vibration energy harvester without being damaged during the fabrication process.

**References**

1. Fang, H.B. et al. *Fabrication and performance of MEMS-based piezoelectric power generator for vibration energy harvesting.* Microelectronics Journal, 2006. 37(11): p. 1280-1284.
2. Liu, J.Q. et al. *A MEMS-based piezoelectric power generator array for vibration energy harvesting.* Microelectronics Journal, 2008. 39(5): p. 802-806.
3. Ferrari, M. et al. *A single-magnet nonlinear piezoelectric converter for enhanced energy harvesting from random vibrations.* Sensors and Actuators A: Physical, 2011. 172(1): p. 287-292.
4. Rezaeisaray, M. et al. *Low frequency piezoelectric energy harvesting at multi vibration mode shapes.* Sensors and Actuators A: Physical, 2015. 228: p. 104-111.
5. Zhou, S. et al. *Analytical and experimental investigation of flexible longitudinal zigzag structures for enhanced multi-directional energy harvesting.* Smart Materials and Structures, 2017. 26(3): p. 035008.
6. Derakhshani, M., Berfield, T.A. Murphy, K.D., *Dynamic Analysis of a Bi-stable Buckled Structure for Vibration Energy Harvester*, in *Dynamic Behavior of Materials, Volume 1*. 2018, Springer. p. 199-208.
7. University of Louisville, Micro/Nano Technology Center. *Website,*. Available from: http://louisville.edu/micronano.
8. Hwang, B.H. et al. *Growth mechanism of reactively sputtered aluminum nitride thin films.* Materials Science and Engineering: A, 2002. 325(1-2): p. 380-388.